\documentclass[11pt]{article}
\usepackage{epsfig}
\usepackage{graphicx}
\usepackage{times,epsfig,makeidx,amsfonts,amsmath}
\usepackage{lscape} 
\usepackage[round]{natbib} 

\usepackage{amsmath}

\def\N{\mbox{N}}

\def\P{\mbox{P}}
\def\d{{\rm d}}

\def\half{\hbox{$1\over2$}}
\def\qrt{\hbox{$1\over4$}}

\begin{document}

\title{On the Equivalence between Bayesian and Classical Hypothesis Testing}
\author{Thomas S. Shively \footnote{McCombs School of Business, University of Texas at Austin. tom.shively@mccombs.utexas.edu}
\and
Stephen G. Walker \footnote{Department of Mathematics \& Division of Statistics and Scientific Computation, University of Texas at Austin. s.g.walker@math.utexas.edu }
}

\date{}
\maketitle

\vspace{0.1in}

\abstract{For hypotheses of the type
$$H_0:\theta=\theta_0\quad \mbox{vs}\quad H_1:\theta\ne\theta_0$$
we demonstrate the equivalence of a Bayesian hypothesis test using a Bayes factor and the corresponding classical test, for a large class of models, which are detailed in the paper. In particular, we show that the role of the prior and critical region for the Bayes factor test is only to specify the type I error. This is their only role since, as we show, the power function of the Bayes factor test coincides exactly with that of the classical test, once the type I error has been fixed. 

For more complex tests involving nuisance parameters, we recover the classical test by using Jeffreys prior on the nuisance parameters, while the prior on the hypothesized parameters can be arbitrary up to a large class. On the other hand, we show that using proper priors on the nuisance parameters results in a test with uniformly lower power than the classical test.  
}

\vspace{0.1in} \noindent Keywords: Bayes factor; Exponential family; Montone function; Power function; Regression model; Uniformly most powerful test.
\vspace{0.2in}

\newpage

\vspace{0.4in}
\noindent
{\bf 1. Introduction}.
There are potentially many ways a Bayesian can select between a specific density model $f(x|\theta_0)$ and the more general model given by $\{f(x|\theta),\,\pi(\theta)\}$, where $\pi(\theta)$ is a prior distribution. However, though there has traditionally been a reluctance for the Bayesian to investigate the decision from a type I error perspective, every decision criterion must have a probability of making the wrong choice, when assuming $f(x|\theta_0)$ to be correct.  For the classes of models we consider in this paper, and when decisions are based on the Bayes factor, we show there is an explicit result connecting the decision criterion and the value of the type I error. Our argument then is that it is preferable for the Bayesian to select the critical region for the Bayes factor using benchmark type I errors. The reasoning is that for any ad-hoc chosen critical region for the Bayes factor, the type I error can be computed and it is unreasonable to allow it to be either too small or too large. Once this type I error has been put in place, we show that the power function for the Bayes factor decision criterion coincides with the power function for the classical test. If the classical test is uniformly most powerful, then we have effectively defined a uniformly most powerful Bayesian test, which differs from the one defined by Johnson (2013).

If the decision criterion to test $H_0: \theta= \theta_0$ vs $H_1:\theta\ne\theta_0$ (or a one-sided alternative $H_1:\theta > \theta_0$) is based on the Bayes factor, i.e.  reject $H_0$ if $B>\lambda$, then, for the models we consider, we show that for any choice of $(\lambda,\pi)$, there exists a $\gamma$ such that
$$B>\lambda\iff T\in C_\gamma,$$
where $T$ is the classical test statistic for the hypothesis test, $C_\gamma$ is a critical region for the test of the form $C_\gamma=\{T: T>\gamma_1\}$ or $C_\gamma=\{T: T>\gamma_1$ or $T<\gamma_2\}$, and $\gamma=\gamma_1$ or $(\gamma_1,\gamma_2)$, depending on the type of testing problem. Hence, the well known problem of selecting  
both $\lambda$ and $\pi$ for the Bayesian is equivalent  to the selection of $\gamma$. In fact, the sole role of $(\lambda,\pi)$ is in determining the type I error; they play no further role in the test. Consequently, we argue that the selection of $\gamma$ based on the value of the type I error is now the most interpretable idea; and certainly makes sense from an Objective Bayesian point of view. This then defines the Bayes factor decision criterion without having to specify a particular prior $\pi$ or value of $\lambda$. 

On the other hand, if the above thinking is eschewed, and a $(\lambda,\pi)$ has been chosen, there still exists a $\gamma$ and the type I error can be evaluated. The power function for the Bayes factor corresponds to the power function of the classical test with the type I error determined by the choice of $(\lambda,\pi)$. Moreover, the test is actually the classical test with a possibly unreasonable type I error.  One is simply working with our recommendations and a classical test except allowing the type I error to be dictated by the choice of $(\lambda,\pi)$ rather than set at a traditional value. 

The key to the paper is working with models $f(x|\theta)$ for which
$$B(t)=\int \frac{f(x|\theta)}{f(x|\theta_0)}\,\pi(\d\theta)=\int g(t|\theta)\,\pi(\d\theta)$$
and $B(t)$ is a monotone (for a one-sided test) or convex (for a two-sided test) function in $t$.
If the alternative hypothesis is one-sided, then for any chosen $\lambda$ there exists a $\gamma_1$ such that
\begin{equation}
\lambda=B(\gamma_1)=\int g(\gamma_1|\theta)\,\pi(\d\theta)\label{key}
\end{equation}
and $B(T)>\lambda$ if and only if $T>\gamma_1$.
If the alternative is two-sided, then for any chosen $\lambda$, we can find a $\pi$ such that there exists $\gamma=(\gamma_1,\gamma_2)$ for which
\begin{equation}
\lambda=B(\gamma_1)=\int g(\gamma_1|\theta)\,\pi(\d\theta)=B(\gamma_2)=\int g(\gamma_2|\theta)\,\pi(\d\theta)\label{key1}
\end{equation}
and $B(T)>\lambda$ if and only if $T<\gamma_1$ or $T>\gamma_2$.

In either case, $B>\lambda$ if and only if $T\in C_\gamma$. We can now set $\gamma$ in a traditional way; i.e.
$$\P_{\theta=\theta_0}(T\in C_\gamma)=\alpha$$
for some standard $\alpha$. This follows since we can find a $\pi$ and a $\lambda$ such that this particular $\gamma$ can be set via (\ref{key}) or (\ref{key1}). And vica versa, the sole role of $(\lambda,\pi)$, as far as the test is concerned, is to determine $\gamma$ and therefore the type I error.

The density functions we consider in this paper are of the form $f(x|\theta,\phi)$ where $\theta$ may be either a scalar or vector, $\phi$ is a nuisance parameter, and the hypothesis test of interest is 
\begin{equation}
H_0:\theta=\theta_0\quad \mbox{vs} \quad H_1:\theta\ne\theta_0. \label{eq16}
\end{equation}
When $\theta$ is a scalar we also consider one-sided tests where the alternative is $H_1:\theta>\theta_0$. When the null hypothesis is a single point $\theta=\theta_0$ and there is no nuisance parameter $\phi$, the Bayes factor is
$$B=\frac{\int f(x|\theta)\,\pi(\d\theta) }{ f(x|\theta_0)}.$$

There is a vast amount of literature on how to select the prior $\pi(\theta)$ for constructing the Bayes factor. It is well known that the choice of prior can significantly influence the value of the Bayes factor. See, for example, Garcia-Donato and Chen (2005). The overwhelming literature is on objective priors for Bayes factors where the goal is to find a default prior that works well across a range of testing problems; see Aitkin (1991) for the posterior Bayes factor,  O'Hagan (1995) for the fractional Bayes factor,  Berger and Perrichi (1996) for the intrinsic Bayes factor, and for other ideas see De Santis and Spezzaferri (1997). 

There is also a significant literature related to the choice of $\lambda$. For example, Jeffreys (1961) gave a scale for determining the evidence in favour of $H_0$. More recently, Kass and Raftery (1995) gave an ad-hoc sliding scale of $\lambda$ values to define the strength of evidence in favor of $H_1$. 

To outline our main result for the one-parameter, one-sided hypothesis test
\begin{equation}
H_0:\theta=\theta_0\quad \mbox{vs} \quad H_1:\theta>\theta_0,  \label{eq22}
\end{equation}
consider a continuous density function $f(x|\theta)$ such that
\begin{equation}
g(t,\theta_1,\theta_2)=\frac{f(x|\theta_2)}{f(x|\theta_1)}  \label{eq23}
\end{equation}
is a monotone increasing function of $t=t(x)$ for every $\theta_0<\theta_1<\theta_2$. The classical test for this problem is to reject $H_0$ if $T>\gamma$ where 
$$\P_{\theta=\theta_0}(T>\gamma)=\alpha$$
for a suitable choice of $\alpha$. This test is a UMP test (see Shi and Tao, 2008, Theorem 3.2.2).

For a Bayesian test using a specific prior $\pi(\theta)$ defined on $\theta>\theta_0$ we show in section 2.1 that there is a unique $\lambda$, see (\ref{key}), such that $B>\lambda$ if and only if $T>\gamma$. Using this value of $\lambda$, the power function for the Bayesian test exactly matches the power function of the classical UMP test. Moreover, we show this result holds for every $\pi(\theta)$ defined on $\theta>\theta_0$ and therefore every Bayesian test, no matter what prior is used, is equivalent to the classical UMP test. Hence, the properties of the Bayesian test are independent of the prior.

We show a similar result in section 2.2 for the two-sided test in (\ref{eq16}) when $f(x|\theta)$ is a member of the one-parameter exponential family
\begin{equation}
f(x_i|\theta)=a(x_i)\exp\{\theta d(x_i)-b(\theta)\} \label{eq7}.
\end{equation}
The classical test for this problem is to reject $H_0$ if $T<\gamma_1$ or $T>\gamma_2$ where $T=\sum_{i=1}^n d(X_i)$, and $\gamma_1$ and $\gamma_2$ are chosen such that $\P_{\theta=\theta_0}(T<\gamma_1)=\P_{\theta=\theta_0}(T>\gamma_2)=\alpha/2.$ This test is a uniformly most powerful unbiased (UMPU) test (see Shi and Tao, 2008, Theorem 3.3.4). 

For a Bayesian test using a prior $\pi\in\Pi$, where $\Pi$ is a large class of prior distributions, we show there is a unique $\lambda$ such that $B>\lambda$ if and only if $T<\gamma_1$ or $T>\gamma_2$. Therefore, the Bayesian test obtained for every $\pi\in\Pi$ is equivalent to the classical UMPU test. For example, if $f(x|\theta)$ is the density function for a $\N(\cdot|\theta,\sigma^2)$ random variable with $\sigma^2$ known, the Bayesian test is equivalent to the classical UMPU test for every symmetric prior centered at $\theta_0$. This means every Gaussian prior with mean $\theta_0$ (no matter what the variance is) will give an equivalent Bayesian test. Other symmetric priors that have been proposed in the literature such as t-distributions and Johnson and Rossell's (2010) non-local method-of-moments distributions also give equivalent UMPU Bayesian tests.

In section 2.3 we consider a two-sided test for the mean of a Gaussian $\N(\cdot|\theta,\sigma^2)$ distribution when $\sigma^2$ is unknown. In this case, $\sigma^2$ is a nuisance parameter and must be integrated out when computing the numerator and denominator of the Bayes factor. We show that if a diffuse prior is used for $\sigma^2$, then the Bayesian test is equivalent to the classical t-test for every symmetric prior centered at $\theta_0$.

Section 3 considers the properties of Bayesian tests in Gaussian regression models. The model we consider is
\begin{equation}
y_i=\sum_{j=1}^p \beta_jx_{ij}+\sigma\epsilon_i \label{eq10}
\end{equation}
where $(\epsilon_i)_{i=1}^n$ are independent standard normal and the test of interest is
\begin{equation}
H_0:\beta=0\quad \mbox{vs} \quad H_1:|\beta|>0 \label{eq19}
\end{equation}
with $|\beta|=\sum_{j=1}^p\beta_j^2.$ 

For $\sigma^2$ known, we show that the Bayesian test is the same for any prior on $\beta$ in the class of elliptical distributions
\begin{equation}
\pi(\beta|\sigma^2)\propto r(\beta'\Sigma^{-1}\beta)  \label{eq11}
\end{equation}
where $\Sigma=\sigma^2 (X'X)^{-1}$ and $X$ is the matrix of regressor variables. Further, we show that every Bayesian test using a prior from this class is equivalent to the classical test for this problem. For $\sigma^2$ unknown, we show that if a diffuse prior is used for $\sigma^2$, then the Bayesian test is equivalent to the classical F-test for every prior in (\ref{eq11}).

Section 4 considers the problem of two-sample tests for the equality of means and variances, and also subset selection for the linear regression model. Here we establish the principle that we recover the classical tests when we place standard diffuse priors on the nuisance parameters while the choice of prior on the hypothesized parameter can be arbitrarily chosen from a large class of prior distributions.
Section 5 then looks at what happens when the Bayesian elects to be informative about all parameter values; both nuisance and those under hypothesis. The result is quite startling in that it can be shown under general conditions that the subjective Bayes factor is uniformly worse than the classical test, or equivalently, uniformly worse than the Bayesian test with diffuse priors for the nuisance parameters. Section 6 considers the implication of the results in sections 2, 3 and 4 regarding how to interpret scales that measure the strength of the evidence of the Bayes factor in favor of the alternative. Section 7 concludes with a discussion.

\vspace{0.2in}
\noindent
{\bf 2. Tests for one-parameter distributions}. This section shows the properties of Bayesian tests for one- and two-sided tests involving one-parameter distributions. Section 2.1 considers one-sided testing problems while section 2.2 discusses two-sided testing problems. Section 2.3 considers a two-sided test of the mean of a Gaussian distribution when $\sigma^2$ is unknown. Section 2.4 discusses the relationship between the results developed in sections 2.1 and 2.2 and a UMP Bayesian test recently proposed by Johnson (2013).

\vspace{0.1in}
\noindent
{\bf 2.1 One-sided tests}. To illustrate the properties of a Bayesian test in a well-known context, consider $(X_i)_{i=1}^n$ from a normal distrbution with unknown mean $\theta$ and known variance $\sigma^2=1$, and a test of (\ref{eq22}) with $\theta_0=0$.

We first consider a simple case where $\pi(\theta)$ is a point prior at $\theta=\theta_1$. Then the appropriate Bayes factor for the test is given by
$$B=\exp\{\theta_1 T - \half n\theta_1^2\}.$$
An important property is that $B=B(T)$ is a monotone increasing function of $T$ where $T=\sum_{i=1}^n X_i$ is the classical test statistic. For any chosen critical value $\lambda$, i.e. the Bayesian rejects $H_0$ if $B>\lambda$, there exists $\gamma=\gamma(\pi,\lambda)$ such that
$$\lambda=\exp\{\theta_1\gamma\ - \half n\theta_1^2\}.$$
Then $B>\lambda$ if and only if $T>\gamma$ and the Bayesian test that rejects $H_0$ if $B>\lambda$ is equivalent to the classical UMP test and is therefore a UMP test itself. It would appear clear now to select $\gamma$ directly using benchmark type I error considerations. If not, the test remains classical but with a possibly unreasonable type I error.

Now consider a general prior $\pi(\theta)$ defined on $\theta>0$. Then the appropriate Bayes factor is given by
\begin{equation}
B=\int_{\theta>0} \exp\{\theta T - \half n\theta^2\}\,\pi(\d\theta)=\int_{\theta>0} g(T,\theta)\,\pi(\d\theta).  \label{eq33}
\end{equation}
where $g(T,\theta)$ is a monotone increasing function of $T$ for any $\theta>0$. Since the integral of an increasing function with respect to any prior $\pi(\theta)$ is also an increasing function, $B$ is an increasing function of $T$. Setting
$$\lambda=\int_{\theta>0} \exp\{\theta \gamma - \half n\theta^2\}\,\pi(\d\theta)$$
implies $B>\lambda$ if and only if $T>\gamma$ for every prior $\pi(\theta)$ defined on $\theta>0$. Therefore, the Bayesian test that rejects $H_0$ if $B>\lambda$ is equivalent to the classical UMP test and is independent of $\pi$. Thus, the Bayesian test is a UMP test no matter what prior is used.

To generalize the Gaussian example, consider the one-sided test of (\ref{eq22}) for a continuous density function $f(x|\theta)$.

\vspace{0.2in}
\noindent
{\sc Theorem 1.} {\sl Let $f(x|\theta)$ be a continuous density function that satisfies (\ref{eq23}). Then the Bayesian test of (\ref{eq22}) that rejects $H_0$ if $B>\lambda$ with $\P_{\theta=\theta_0}(B>\lambda)=\alpha$ is independent of the prior $\pi$ and is a UMP test.}

\vspace{0.2in}
\noindent
{\sc Proof.} As discussed in the introduction, the classical UMP test for this problem rejects $H_0$ if $T>\gamma$. The Bayes factor for this test is
$$B=\int_{\theta>\theta_0}g(T,\theta_0,\theta)\pi(\d\theta)$$
and is a monotone increasing function of $T$ for every prior $\pi(\theta)$. Then, setting
$$\lambda=\int_{\theta>\theta_0}g(\gamma,\theta_0,\theta)\pi(\d\theta)$$
we have $B>\lambda$ if and only if $T>\gamma$ for any prior $\pi(\theta)$. Therefore, the Bayesian test that rejects $H_0$ if $B>\lambda$ is equivalent to the classical UMP test for every prior $\pi(\theta)$, and therefore every Bayesian test is a UMP test. \hfill $\square$

\vspace{0.1in}
\noindent
We note that continuous density functions in the exponential family are members of this class because
$$g(t,\theta_1,\theta_2)=\frac{f(x|\theta_2)}{f(x|\theta_1)}=\exp\{d(x)(\theta_2-\theta_1)-n[b(\theta_2)+b(\theta_1)] \}$$
is an increasing function of $t(x)=\sum_{i=1}^n d(x_i)$ for $\theta_2>\theta_1>\theta_0$. Also, this result can be generalized to discrete distributions in a straightforward manner, although the notation becomes more cumbersome due to the need to randomize to get an exact $\alpha$-level test.

\vspace{0.1in}
\noindent
{\bf 2.2 Two-sided tests}. We now consider the two-sided test in (\ref{eq16}). To illustrate the properties of a two-sided Bayesian test in a well-known context, we again consider $(X_i)_{i=1}^n$ from a normal distrbution with unknown mean $\theta$,  known variance $\sigma^2=1$, and $\theta_0=0$. The classical test for this problem is to reject $H_0$ if $T>\gamma$ or $T<-\gamma$ where $T=\sum_{i=1}^n X_i$ and $\gamma$ is chosen so that $\P_{\theta=0}(T>\gamma)=\alpha/2$. This is a uniformly most powerful unbiased (UMPU) test.

For a symmetric prior $\pi(\theta)$ centered at 0 the appropriate Bayes factor is given by
$$B=\frac{\int_{\theta>0}[f(x|\theta)+f(x|-\theta)]\pi(\d\theta)}{f(x|\theta=0)}$$
$$=\int_{\theta>0}[\exp\{\theta T\}+\exp\{-\theta T\}]\exp\{-\half n\theta^2\}\pi(\d\theta)$$
$$=\int_{\theta>0}h(T,\theta)\pi(\d\theta)$$
where $h(T,\theta)$ is a convex function of $T$ for any $\theta$. Also, if
$$h(\gamma,\theta)=[\exp\{\theta \gamma\}+\exp\{-\theta \gamma\}]\exp\{-\half n\theta^2\}$$
then $h(T,\theta)>h(\gamma,\theta)$ if and only if $T>\gamma$ or $T<-\gamma$. Since this is true for every $\theta>0$, if we set
$$\lambda=\int_{\theta>0}h(\gamma,\theta)\pi(\d\theta),$$
then $B>\lambda$ if and only if $T>\gamma$ or $T<-\gamma$. Therefore, the Bayesian test that rejects $H_0$ if $B>\lambda$ is equivalent to the classical UMPU test and is independent of the choice of prior from the class of all symmetric distributions centered at zero. Since the classical and Bayesian tests are equivalent, the Bayesian test is a UMPU test for any symmetric prior centered at zero.

This result can be generalized to density functions in the exponential fammily. More specifically, if $f(x|\theta)$ is a continuous density function in the exponential family of density functions described in (\ref{eq7}) then there exists a class of prior distributions $\Pi$ defined on the support of $\theta$ such that the Bayesian test of (\ref{eq16}) that rejects $H_0$ if $B>\lambda$ with $P_{\theta=\theta_0}(B>\lambda)=\alpha$ is independent of the prior $\pi\in\Pi$ and is a UMPU test.

As discussed in the introduction, the classical UMPU test for this problem is to reject $H_0$ if $T<\gamma_1$ or $T>\gamma_2$ where $T=\sum_{i=1}^n d(X_i)$.

To construct the class of prior distributions $\Pi$ and compute $\lambda$ so that the Bayesian test is equivalent to the classical UMPU test, let
$$h(t,\theta_1,\theta_2)=\frac{f(x|\theta_1)}{f(x|\theta_0)} + \frac{f(x|\theta_2)}{f(x|\theta_0)}$$
where $\theta_2<\theta_0<\theta_1$ and $t=\sum_{i=1}^n d(x_i)$. Then $h(t,\theta_1,\theta_2)$ is a convex function of $t$. Further, for every $\theta>\theta_0$ there exists a unique $\tilde{\theta}=r(\theta)<\theta_0$ such that
$$h(\gamma_1,\theta,r(\theta))=h(\gamma_2,\theta,r(\theta)).$$
Now let $\Pi$ be the class of prior distributions such that for $\pi\in\Pi$ we have $\pi(\theta)=\pi(r(\theta))$ for all $\theta>\theta_0$. For the Gaussian case discussed above with $\theta_0=0$, $r(\theta)=-\theta$ and $\Pi$ is the class of symmetric prior distributions centered at 0. 

The appropriate Bayes factor for this problem is
$$B=\int_{\tilde{\theta}<\theta_0}\frac{f(x|\tilde{\theta})}{f(x|\theta_0)}\pi(\tilde{\theta}) \d\tilde{\theta}+\int_{\theta>\theta_0}\frac{f(x|\theta)}{f(x|\theta_0)}\pi(\theta) \d\theta$$
$$=\int_{\theta>\theta_0}h(t,\theta,r(\theta))\pi(\theta) \d\theta.$$
If we set
$$\lambda=\int_{\theta>\theta_0}h(\gamma_1,\theta,r(\theta))\pi(\theta) \d\theta$$
then $B>\lambda$ if and only if $T>\gamma_1$ or $T<\gamma_2$ for any $\pi\in\Pi$. Therefore, the Bayesian test that rejects $H_0$ if $B>\lambda$ is equivalent to the classical UMPU test for every prior $\pi\in\Pi$, and therefore every Bayesian test using one of these priors is a UMPU test. 

This result is formalized in the following theorem:

\vspace{0.1in}
\noindent
{\sc Theorem 2.} {\sl Suppose
$$B(t)=\int h(t,\theta)\,\pi(\d\theta)$$
where 
$$h(t,\theta)=\frac{f(x|\theta)}{f(x|\theta_0)}.$$
Assume $B(t)$ is convex, which it is when we have the exponential family. Then for $\gamma_1$ and $\gamma_2$ as defined, with $\gamma_1<\gamma_2$, choose the prior $\pi(\theta)$ so that
$$B(\gamma_1)=B(\gamma_2).$$
Then due to the convexity of $B(t)$, it follows that $B(t)>\lambda=B(\gamma_1)=B(\gamma_2)$ if and only if $t<\gamma_1$ or $t>\gamma_2$.}

\vspace{0.1in}
\noindent
Hence, the two-sided test imposes a constraint on the prior which is not present for the one-sided test. However, this constaint is minimal, being effectively a symmetry condition.

\vspace{0.1in}
\noindent
{\bf 2.3 Tests for Gaussian models with $\sigma^2$ unknown}. This section considers two-sided Bayesian tests of (\ref{eq16}) when $(X_i)_{i=1}^n$ are from a normal distrbution with unknown mean $\theta$ and unknown variance $1/\phi=\sigma^2$. In this problem, $\sigma^2$ is a nuisance parameter. The classical test rejects $H_0$ if $T<-\gamma$ or $T>\gamma$ where $T=\sqrt{n}(\bar{X}-\theta_0)/S_X$ with $S_X^2=\frac{1}{n-1}\sum_{i=1}^n (X_i-\bar{X})^2$. We assume $\theta_0=0$. We use the standard diffuse prior for $\phi$ for reasons expanded on in section 5.

\vspace{0.1in}
\noindent
{\sc Lemma 1.} {\sl With prior distributions 
\begin{equation}
\pi(\theta|\phi)=h(\theta\sqrt{\phi})\sqrt{\phi}\quad\mbox{and}\quad\pi(\phi)\propto \phi^{-1},  \label{eq34}
\end{equation}
where $h(\cdot)$ is a symmetric density function centered at 0,
the Bayes factor, given by
\begin{equation}
B=\frac{\int\int f(x|\theta,\phi) \pi(\d\theta|\phi)\pi(\d\phi)} {\int f(x|\theta_0,\phi) \pi(\d\phi)}, \label{eq26}
\end{equation}
is a monotone function in $T^2$.}

\vspace{0.2in}
\noindent
{\sc Proof.}
The denominator of (\ref{eq26}) is given by, and we only consider the relevant terms,
$$\left(\sum_{i=1}^n X_i^2\right)^{-n/2}.$$
The numerator, again only including relevant terms, is given, after some initial transformation $\theta=s/\sqrt{\phi}$, by
$$\int \phi^{n/2-1}\exp\left\{-\half\sum_{i=1}^n X_i^2\phi\right\}\int_0^\infty 2\cosh(n\bar{X}\sqrt{\phi}s)\,h^*(s)\,\d s\,\d\phi,$$
where $h^*(s)=\exp\{-\half ns^2\}\,h(s)$ is a symmetric function. Hence, since $\cosh(\cdot)$ is a symmetric non-negative function, we can write, for positive $(a_j)$,
$$\int_0^\infty 2\cosh(n\bar{X}\sqrt{\phi}s)\,h^*(s)\,\d s=\sum_{j=0}^\infty a_j (n^2\bar{X}^2\phi)^j.$$
Therefore, the Bayes factor is given by
$$B =\kappa\, \sum_{j=0}^\infty \tilde{a}_j\,\bar{X}^{2j} \frac{\Gamma(n/2+j)\,\left(\sum_{i=1}^n X_i^2\right)^{n/2} }
{\left(\sum_{i=1}^n X_i^2\right)^{n/2+j} }$$
$$=\kappa\,\sum_{j=0}^\infty a_j^* \left(\frac{\bar{X}^2}{\sum_{i=1}^n X_i^2}\right)^j,$$
where $\kappa$ does not depend on the data.
The term 
$$\frac{\bar{X}^2}{\sum_{i=1}^n X_i^2}=\frac{1}{n}\,\frac{T^2}{(n-1)+T^2}$$
is a monotone increasing function in $T^2$. \hfill $\square$

\vspace{0.1in}
\noindent
Hence, the Bayes factor is an increasing function of $T^2$. Therefore, there is a unique $\lambda$ that is a function of $\gamma$ such that $B>\lambda$ if and only if $T<-\gamma$ or $T>\gamma$, and the Bayesian test is equivalent to the classical t-test for any symmetric prior $\pi(\theta|\phi)$ defined in (\ref{eq34}).

\vspace{0.1in}
\noindent
{\bf 2.4 Optimal Bayesian tests in the literature}. Recently, Johnson (2013) proposed a definition of a UMP Bayesian test based on finding the prior $\pi(\theta)$ for which
$$\mbox{P}_{\theta}(B>\lambda)\geq \mbox{P}_{\theta}(B'>\lambda)$$
for all $\theta$ and for all
$$B'=\frac{\int f(x|\theta)\,\pi'(\d\theta)}{f(x|\theta_0)},$$
where $\pi'$ is any prior distribution.

To facilitate a comparison with the results developed in sections 2.1 and 2.2, it is convenient to illustrate this idea for the exponential family distribution
$$f(x|\theta)=c(x)\,\exp\{x\theta-b(\theta)\}$$
where $b(\cdot)$ is increasing, and a test of (\ref{eq22}). First, define
$$g_\lambda(\theta,\theta_0)=\frac{\log\lambda+n(b(\theta)-b(\theta_0)) }{\theta-\theta_0}$$
and let $\theta^*$ be the minimizer of $g_\lambda(\theta,\theta_0)$ (assuming for convenience it is unique). Then the UMP Bayesian test Johnson (2013) proposes is to let the prior $\pi$ be a point mass at $\theta^*$ and reject $H_0$ if $B>\lambda=g_\lambda(\theta^*,\theta_0)$.

A disadvantage of this test is that there is no notion of setting the decision criterion $\lambda$ to give a specific type I error. To fairly compare Bayesian and classical tests it is important to control for the type I error rate. Otherwise, the power function can be made arbitrarily close to one for any value of $\theta$ by allowing a sufficiently high probability of type I error.

It is also useful to note that the results in section 2.1 show every prior $\pi$ gives a UMP test of (\ref{eq22}), including the prior with a point mass at $\theta^*$, if $\lambda$ is chosen so that $\P_{\theta=\theta_0}(B>\lambda)=\alpha$.

\vspace{0.2in}
\noindent
{\bf 3. Tests involving regression models.}
In this section we consider the Bayes factor for the Gaussian regression model in (\ref{eq10}) and tests of (\ref{eq19}). Section 3.1 discusses the case where $\sigma^2$ is known while section 3.2 considers the case where $\sigma^2$ is unknown.

The majority of existing research in this area is in the design of a suitable prior distribution for the non-null models and many types of priors have been proposed. Examples include the intrinsic prior of Berger and Perrichi (1996), the mixtures of $g$-priors, see Liang et al. (2008), and Johnson and Rossell's (2010) non-local method-of-moment multivariate priors. Bayarri et al. (2012) contains a thorough discussion of the use of objective priors for this problem. The consistency of some of the resulting Bayes factors is provided in Casella et al. (2009). 

\vspace{0.1in}
\noindent
{\bf 3.1 Tests for Gaussian regression models with $\sigma^2$ known}. Letting $\sigma^2=1$, the model in (\ref{eq10}) can be written as
\begin{equation}
y=X\beta+\epsilon \label{eq12}
\end{equation}
where $y=(y_1,\ldots,y_n)'$, $\beta=(\beta_1,\ldots,\beta_p)'$, $\epsilon=(\epsilon_1,\ldots,\epsilon_n)'$, and $X$ is the $n\times p$ design matrix. Rather than work directly with (\ref{eq12}) we consider the transformed model
$$y=Z\delta+\epsilon$$
where $Z=XQ$, $Z'Z=I$ and $\delta=Q^{-1}\beta$. The equivalent hypothesis test of interest is
\begin{equation}
H_0:\delta=0\quad \mbox{vs} \quad H_1:|\delta|>0 \label{eq13}
\end{equation}
where $|\delta|=\sum_{j=1}^p\delta_j^2$ and the transformed prior for $\delta$ is in the class of spherically symmetric distributions
$$\pi(\delta)\propto r(\delta'\delta).$$

The classical test for this problem is to reject $H_0$ if $|T|=\sum_{j=1}^p T_j^2>\gamma$ where
$$T_j=\sum_{i=1}^n Y_iz_{ij}$$
for $j=1,\ldots,p$, and $\gamma$ is chosen so that $P_{\delta=0}(|T|>\gamma)=\alpha$. This is a likelihood ratio test and is the analog to the well-known F-test when $\sigma^2$ is known.

The appropriate Bayes factor is given by
$$B(T_1,\ldots,T_p)=\int\ldots\int \exp\left\{\sum_{j=1}^p\delta_jT_j\right\}\exp\left\{-\half n|\delta| \right\} \,\pi(|\delta|)\,\d\delta.$$
If $\pi(\delta)$ is a spherically symmetric distribution then $\pi(\d\delta)=\pi(|\delta|)\,\d\delta_1\ldots\d\delta_p$. Hence, 
$$B(T_1,\ldots,T_p)=\int\ldots\int \exp\left\{\sum_{j=1}^p\delta_jT_j\right\}\,g(|\delta|)\,\d\delta$$
where 
$$g(|\delta|)=\exp\left(-\half n|\delta| \right)\pi(|\delta|).$$

\vspace{0.2in}
\noindent
{\sc Theorem 3.} {\sl It is that 
$$B(T_1,\ldots,T_p)=\psi(|T|)$$
where $\psi$ is a montone increasing function.} 

\vspace{0.2in}
\noindent
{\sc Proof.}
Now, for $j=1,\ldots,p$,
$$\partial B/\partial T_j=\int\ldots\int \delta_j\exp\left\{\sum_{j=1}^p\delta_jT_j\right\}\,g(|\delta|)\,\d\delta.$$
Using integration by parts, with
$$u=\exp\left\{\sum_{j=1}^p\delta_jT_j\right\}\quad\mbox{and}\quad v'=\delta_j\,g(|\delta|)$$
we have
$$\partial B/\partial T_j=-T_j\int\ldots\int \exp\left\{\sum_{j=1}^p\delta_jT_j\right\}\,G(|\delta|)\,\d\delta,$$
where $G'=g$.

Letting
$$B'=\int\ldots\int \exp\left\{\sum_{j=1}^p\delta_jT_j\right\}\,G(|\delta|)\,\d\delta.$$
gives the partial differential equations
$$\partial B/\partial T_j=-T_j B'$$
for $j=1,\ldots,p$. The general solution to these equations is of the type
$$B(T_1,\ldots,T_p)=\psi(|T|).$$
But we know that 
$$B(T_1,0,\ldots,0)=\psi(T_1^2)$$
and that $\psi$ must be monotone. In fact, it is easy to show that, for some constant $c>0$, we have
$$\psi(s)=c\,\int_{0}^\infty \cosh (\delta s)\,g(|\delta|)\,\d\delta $$
which is an increasing function for $s>0$ because $\cosh(s)$ is an increasing function for $s>0$. \hfill $\square$

\vspace{0.1in}
\noindent
Therefore, $B(T_1,\ldots,T_p)$ is a monotone increasing function of $|T|$ and $$B(T_1,\ldots,T_p)>\lambda(\gamma)$$ if and only if $|T|>\gamma$ where
$$\lambda(\gamma)=\psi(\gamma).$$
This implies the Bayesian test that rejects $H_0$ if $B(T_1,\ldots,T_p)>\lambda(\gamma)$ is equivalent to the classical test and is independent of the choice of prior from the class of all spherically symmetric priors centered at 0.

\vspace{0.1in}
\noindent
{\bf 3.2 Gaussian regression models with $\sigma^2$ unknown}. This section considers tests of (\ref{eq13}) when $\sigma^2$ is unknown and must be integrated out of the Bayes factor. The classical test for this problem is to reject $H_0$ if
$$F= \frac{(\mbox{RSS}_1-\mbox{RSS}_2)/p }{\mbox{RSS}_2/(n-p)  } >\gamma$$
where $\gamma$ is chosen so $\P_{\delta=0}(F>\gamma)=\alpha$. This is the well-known F-test.
Here
$$\mbox{RSS}_1=y'y$$
and
$$\mbox{RSS}_2=y'(I-H)y$$
where 
$$H=Z(Z'Z)^{-1}Z'=ZZ'$$
is the usual hat matrix. Hence, the F-test involves the statistic
$$T=y'Hy/y'y.$$
In fact, 
$$F=\kappa\,T/(1-T)$$
which is increasing in $T$ and $\kappa$ is a constant not involving $T$.

We now show we can recover the $F$ test with a spherically symmetric prior for $\delta$ and the usual noninformative prior for $\phi=\sigma^{-2}$. 

\vspace{0.1in}
\noindent
{\sc Lemma 2.} {\sl Using the priors $\pi(\delta|\phi)=\phi^{p/2}h(\sqrt{\phi}\delta)$ and $\pi(\phi )\propto \phi^{-1}$, the Bayes factor is a monotone function in $F$.}

\vspace{0.1in}
\noindent
{\sc Proof.}
The appropriate Bayes factor for the test is given by
$$B=\frac{\int\int f(y|\delta,\phi) \pi(\d\delta|\phi)\pi(\d\phi)} {\int f(y|\delta=0,\phi) \pi(\d\phi)}.$$
Following the same reasoning as in section 2.3, the numerator of the Bayes factor, including only relevant terms, is given by
$$\int \phi^{n/2-1}\exp\left\{-\half\phi \,y'y \right\}\sum_{j=0}^\infty a_j\,\phi^j (y'ZZ'y)^j\,\d\phi.$$
This becomes
$$\sum_{j=0}^\infty a_j\,(y'Hy)^j \frac{\Gamma(n/2+j)}{y'y^{n/2+j}}.$$
The denominator of the Bayes factor, again only including relevant terms, is given by
$y'y^{-n/2}$ and hence the Bayes factor can be written, for some $\kappa'$ not depending on the data, as
$$B=\kappa'\,\sum_{j=0}^\infty a_j^*\,\left(\frac{y'Hy}{y'y}\right)^{j}=\kappa'\,\sum_{j=0}^\infty a_j^*\,T^j=\kappa'\,\sum_{j=0}^\infty a_j^* \left(F/(\kappa+F)\right)^j.$$
This is an increasing function of $F$. \hfill $\square$

\vspace{0.1in}
\noindent
If we now set
$$\lambda=\kappa'\,\sum_{j=0}^\infty a_j^* \left(\gamma/(\kappa+\gamma)\right)^j$$
then $B>\lambda$ if and only if $F>\gamma$. This implies the Bayesian test is equivalent to the classical F-test and is independent of the choice of prior for $\delta$ from the class of all spherically symmetric priors centered at 0.

\vspace{0.2in}
\noindent
{\bf 4. Two-sample tests and subset selection.} In this section we consider Bayesian tests based on samples from two Gaussian distributions. Section 4.1 considers the test for the equality of the two means assuming that the variances are known. Section 4.2 considers the two sample $t$-test in which the variances are unknown but equal and section 4.3 considers the equality of variance $F$-test. Section 4.4 looks at subset selection for the linear regression model.

\vspace{0.1in}
\noindent
{\bf 4.1 Tests for the equality of means with known variances.} Here we consider $(X_{i1})_{i=1}^{n_1}$ from a normal distribution with unknown mean $\theta_1$ and known variance $1/\tau_1$, $(X_{i2})_{i=1}^{n_2}$ from a normal distribution with unknown mean $\theta_2$ and known variance $1/\tau_2$, and a test of 
$$H_0:\theta_1=\theta_2\quad\mbox{vs}\quad H_1:\theta_1\ne\theta_2.$$
The classical test for this problem is to reject $H_0$ if $T>\gamma$ where $T=(\bar{X}_1-\bar{X}_2)^2$, $\bar{X}_1$ and $\bar{X}_2$ are the sample means, and $\gamma$ is chosen so that $P_{\theta_1=\theta_2}(T>\gamma)=\alpha$. 

Using the prior $\theta_j$ $\sim$ $\N(0,(cn_j\tau_j)^{-1})$ for $j=1,2$ for some fixed $c>0$ and the prior $\N(0,(cn_1\tau_1+cn_2\tau_2)^{-1}))$ for the common mean under $H_0$, we show that the Bayes factor test does not depend on $c$ and is equivalent to the classical test.

The appropriate Bayes factor is given by
$$B=\frac{\prod_{j=1}^2\int \prod_{i=1}^{n_j}\N(x_{ij}|\theta_j,\tau_j^{-1})\,\N(\d\theta_j|0,(cn_j\tau_j)^{-1})}{\int \prod_{i=1}^{n_1}\N(x_{i1}|\theta,\tau_1^{-1})\,\prod_{i=1}^{n_2}\N(x_{i2}|\theta,\tau_2^{-1})\,\N(\d\theta|0,(cn_1\tau_1+cn_2\tau_2)^{-1})}.$$
Now the terms
$$\exp\left\{-\half \tau_1\sum_{i=1}^{n_1}x_{i1}^2-\half \tau_2\sum_{i=1}^{n_2}x_{i2}^2\right\}$$
cancel from the numerator and denominator and
so, for some $\kappa>0$ not depending on the data, we have 
$$B=\kappa\,\exp\left\{ -\half\frac{1}{1+c}\left[\frac{(n_1\tau_1\bar{x}_1+n_2\tau_2\bar{x}_2)^2}{n_1\tau_1+n_2\tau_2}-n_1\tau_1\bar{x}_1^2- n_2\tau_2\bar{x}_2^2\right]\right\}.$$
We then deduce using straightforward algebra that
$$B=\kappa\,\exp\{\kappa'\,(\bar{x}_1-\bar{x}_2)^2\}$$
where $\kappa'>0$ does not depend on the data. Therefore, if we set $\lambda=\kappa\,\exp\{\kappa'\,\gamma\}$, then $B>\lambda$ if and only if $T>\gamma$. Hence, the Bayesian test that rejects $H_0$ if $B>\lambda$ is equivalent to the classical test for all $c$.

\vspace{0.1in}
\noindent
{\bf 4.2 Tests for the equality of means with equal but unknown variances.}
We now consider the case where the variances are unknown, but equal; so let $\phi=\tau_1=\tau_2$.
Since the variances are equal we can re-parameteize the $\theta$s. Therefore, $(X_{i1})_{i=1}^{n_1}$ come from a normal distribution with unknown mean $\theta_1$ and unknown variance $1/\phi$, $(X_{i2})_{i=1}^{n_2}$ from a normal distribution with unknown mean $\theta_1+\theta$ and unknown variance $1/\phi$, and we are interested in a test of 
$$H_0:\theta=0\quad\mbox{vs}\quad H_1:\theta\ne0.$$
The classical test for this problem is to reject $H_0$ if $T>\gamma$ or $T<-\gamma$ where 
$$T=\frac{\bar{X}_2-\bar{X}_1}{\sqrt{(n_1-1)S_1^2+(n_2-1)S_2^2}}$$
and $\gamma$ is chosen so that $P_{\theta=0}(T>\gamma)=\alpha/2$. 

We adopt standard non-informative priors for the nuisance parameters, namely
$$\pi(\theta_1,\phi)\propto \phi^{-\half}. $$
The prior for $\theta$ is normal with zero mean and variance $(c\phi)^{-1}$ and the aim is to show that the Bayes factor test does not depend on $c$.

The Bayes factor is, after the necessary integration, given by
$$B=\kappa\,\left(\frac{(n-1)S^2}
{(n-1)S^2-\frac{n_2^2(\bar{x}_2-\bar{x})^2}{c+n_2-n_2^2/n}}
\right)^{n/2}$$
where $\kappa>0$ is a constant not depending on the data and $S^2$ is the sample variance of the whole data set. Now
$$\bar{x}_2-\bar{x}=\frac{n_1}{n}(\bar{x}_2-\bar{x}_1)$$
and hence
$$B=\kappa\left(\frac{1}{1-\kappa' \tilde{T}^2}\right)^{n/2}$$
where
$$\tilde{T}^2=\frac{(\bar{X}_2-\bar{X}_1)^2}{(n-1)S^2}$$
and $\kappa'>0$ does not depend on the data. Using
$$(n-1)S^2=(n_1-1)S_1^2+(n_2-1)S_2^2+\frac{n_1n_2}{n}(\bar{X}_2-\bar{X}_1)^2, $$
where $S_1^2$ and $S_2^2$ are the sample variances from the $(X_{i1})$ and $(X_{i2})$ samples, respectively, 
we see that
$$\tilde{T}^2=\frac{T^2}{1+T^2\,n_1n_2/n}$$
where $T$ is the classical test statistic.

Finally, $\tilde{T}^2$ is increasing with $T^2$, since $n_1,n_2>1$ implies $n_1n_2\geq n$, and $B$ is increasing with $\tilde{T}^2$. Therefore, $B$ is a monotone function in $T^2$ which means we can recover the classical two-sample $t$-test for all $c>0$ by taking the appropriate $\lambda$.

\vspace{0.1in}
\noindent
{\bf 4.3 Test for equality of two variances.} In this case we assume $(X_{i1})_{i=1}^{n_1}$ come from a normal distribution with unknown mean $\mu_1$ and unknown variance $\phi^{-1}$, and $(X_{i2})_{i=1}^{n_2}$ come from a normal distribution with unknown mean $\mu_2$ and unknown variance $(\theta\phi)^{-1}$. We are interested in a test of 
$$H_0:\theta=1\quad\mbox{vs}\quad H_1:\theta>1.$$
The classical F-test for this problem is to reject $H_0$ if $F>\gamma$ where 
$$F=S_1^2/S_2^2$$
with
$$\quad\,S_1^2=\sum_{i=1}^{n_1}(X_{i1}-\bar{X}_1)^2\quad\mbox{and}\quad S_2^2=\sum_{i=1}^{n_2}(X_{i2}-\bar{X}_2)^2$$
and $\gamma$ is chosen so that $P_{\theta=1}(F>\gamma)=\alpha$.

The prior for $\theta$ will be denoted by $\pi(\theta)$. The priors for the nuisance parameters will be diffuse, so the prior for $\phi$ is proportional to $\phi^{-1}$, and the prior for the $\mu_j$ will be proportional to 1. The Bayes factor is then given by
$$B=\kappa\,\,\frac{\int\int \phi^{n/2-1}\theta^{n_2/2}\exp\{-\half\phi (S_1^2+\theta S_2^2)\} \,\d\phi\,\pi(\d\theta) }
{\int \phi^{n/2-1} \exp\{-\half\phi (S_1^2+S_2^2)\}\,\d\phi}.$$
where $\kappa$ does not depend on the data. This leads to
$$B=\kappa\int_{\theta>1} \theta^{n_2/2}\left(\frac{S_1^2+S_2^2}{S_1^2+\theta S_2^2}\right)^{n/2}\,\pi(\d\theta)$$
and hence
$$B=\kappa \int_{\theta>1} \theta^{n_2/2}\left(\frac{F+1}{F+\theta}\right)^{n/2}\,\pi(\d\theta)$$
is monotone increasing in $F$. Therefore, if we set 
$$\lambda=\kappa\int_{\theta>1}\theta^{n_2/2}\left(\frac{\gamma+1}{\gamma+\theta}\right)^{n/2}\,\pi(\d\theta)$$ 
then $B>\lambda$ if and only if $F>\gamma$ and the Bayesian test that rejects $H_0$ if $B>\lambda$ is equivalent to the classical F-test. We can deal with a two-sided test by following the work found in section 2.

\vspace{0.1in}
\noindent
{\bf 4.4 Subset selection.} Here we revisit the linear regression model but inspired now with the knowledge that using non-informative priors on the nuisance parameters leads to the classical tests.
 
Consider the linear model
$$y=X_1\beta_1+X_2\beta_2+\sigma\varepsilon$$
where $X_1$ is $n\times p_1$, $X_2$ is $n\times p_2$, $\varepsilon$ is normal with zero mean and variance-covariance matrix the $I_n$ identity matrix, and a test of the hypothesis
$H_0:\beta_2=0$ vs. $H_1:\beta_2\ne 0.$ The classical F-test rejects $H_0$ if
$$F=\frac{y'(H-H_1)y}{y'(I-H)y}>\gamma,$$
where $H_1$ is the hat matrix with $\beta_2=0$ and $H$ is the full hat matrix, 
and where $P_{\beta_2=0}(F>\gamma)=\alpha$.

Letting $\phi=\sigma^{-2}$, we take the prior for $\beta_2|\phi$ as
$$\pi(\beta_2|\phi)\propto \exp\left\{-\half c \phi \beta_2'X'X\beta_2 \right\}$$
for some $c>0$. Also, we adopt the standard non-informative priors for the nuisance parameters so $\pi(\beta_1,\phi)\propto \phi^{-1}$. Our aim is to show that the Bayes factor test does not depend on $c$ and is equivalent to the F-test. 

Using these priors, the denominator of the Bayes factor is, retaining only relevant terms, given by
$$\bigg( y'(I-H_1)y  \bigg)^{-n/2}$$
where 
$$H_1=X_1(X_1'X_1)^{-1}X_1'.$$
For the numerator, let us define
$$X=(I-H_1)X_2.$$
After the necessary integration, it is possible to show that the numerator is, again with only relevant terms, given by
$$\bigg( y'(I-H_1)y-\frac{1}{1+c}y'X (X'X)^{-1}X'y \bigg)^{-n/2}.$$
Hence, the Bayes factor test statistic is a monotone function of
$$\frac{y'(I-H_1)y}{y'(I-H_1)y-\frac{1}{1+c} y'X(X'X)^{-1}X'y},$$
and  therefore a monotone function of
$$T=\frac{y'X(X'X)^{-1}X'y}{y'(I-H_1)y}.$$
If we define $\tilde{X}=[X_1 X_2]$, then
it is easy to show that
$$H_1+X(X'X)^{-1}X'=\tilde{X}(\tilde{X}'\tilde{X})^{-1}\tilde{X}'$$
and hence 
$$T=\frac{F}{1+F},$$
which is monotone in $F$. Therefore, the Bayes factor test is equivalent to the classical F-test for all $c$.

\vspace{0.1in}
\noindent
{\bf 5. Subjective Bayes factor.} For the testing problems considered in sections 2, 3 and 4 we showed that if standard non-informative diffuse priors are used for nuisance parameters then we recover the classical test for any prior on the parameter under hypothesis chosen from a wide class of distributions. In many cases, the resulting tests are UMP or UMPU tests. However, if we alter this and instead put proper priors on the nuisance parameters, we show in Theorem 4 in this section that the resulting subjective Bayes factor test is uniformly worse than the classical test.

We begin by illustrating the result for the well-known two-sided equality of variance test. Consider two models where $(X_{i1})_{i=1}^{n_1}$ come from a normal distribution with known mean 0 and unknown variance $\tau_1^{-1}$ and $(X_{i2})_{i=1}^{n_1}$ come from a normal distribution with known mean 0 and unknown variance $\tau_2^{-1}$. 

The classical test for this problem is to reject $H_0$ if $F<\gamma_1$ or $F>\gamma_2$ where 
$$F=S_1^2/S_2^2\quad\mbox{with}\quad S_j^2=\sum_{i=1}^{n_j}X_{ij}^2$$ 
and $\gamma_1$ and $\gamma_2$ are chosen so the probability of a Type I error is $\alpha$. This is the well-known F-test.

To keep the notation manageable in our illustration, suppose the informative priors for the $\tau_j$ are independent $\mbox{Gamma}(a_j,b_j)$ distributions, the informative prior for the common variance is a $\mbox{Gamma}(a,b)$ distribution, and set $n_1=n_2=n/2$, $a_1=a_2=a/2$ and $b_1=b_2=b/2$. Then the appropriate Bayes factor is given by
$$B=\kappa\,\,\frac{(b/S^2+\half)^{a+n/2}}{(\frac{1}{2}b/S^2+\half \tilde{T})^{a/2+n/4} \,(\frac{1}{2}b/S^2+\half(1-\tilde{T}))^{a/2+n/4}},$$
for some $\kappa>0$, which does not depend on the $(X_{ij})$, $S^2=S_1^2+S_2^2$ and $\tilde{T}=S_1^2/S^2$.

Following some extensive algebra and removing terms that do not depend on the data, the subjective Bayes factor test statistic is given by
$$B^*(Q,T)=\frac{Q+\half}{\sqrt{(Q+\half)^2-T}}$$
where $Q=b/S^2$ and $T=\qrt-\tilde{T}(1-\tilde{T})$. Hence,
$$T=\qrt-\frac{F}{(1+F)^2}$$ and we pick $\gamma_1<\gamma_2$ such that
$$\frac{\gamma_1}{(1+\gamma_1)^2}=\frac{\gamma_2}{(1+\gamma_2)^2}$$
and 
$$\gamma=\qrt-\frac{\gamma_1}{(1+\gamma_1)^2}$$
so $F<\gamma_1$ or $F>\gamma_2$ if and only if $T>\gamma$.

The conditions of Theorem 4 require that $B^*(Q,T) \le B^*(0,T)$ for all $T$, and $B^*(Q,T)$ is monotone increasing in $T$ for all $Q$. This is easily shown in this testing problem. Therefore, given these conditions, Theorem 4 shows that the subjective Bayes factor test has uniformly lower power than the classical test, or equivalently, lower power than the Bayes factor test with a diffuse prior on the nuisance parameter.

We now state and prove Theorem 4. Consider a test of $H_0:\theta=\theta_0$ vs. $H_1:\theta \ne \theta_0$ where $H_0$ is rejected if $\psi(Q,T)>\lambda$. Further, suppose $\psi(Q,T) \le \psi(0,T)$ for all $T$, and $\psi(Q,T)$ is monotone increasing in $T$ for all $Q$. Under these conditions we can prove the following:

\vspace{0.1in}
\noindent
{\sc Theorem 4.} {\sl If 
$$\alpha=\P_{\theta_0}(\psi(Q,T)>\lambda)=\P_{\theta_0}(T>\gamma)$$
then for all $\theta$ it is that
$$P_\theta (\psi(Q,T)>\lambda)\leq \P_\theta(T>\gamma).$$
That is, the test based on $\psi(Q,T)>\lambda$ is uniformly worse than the test based on $T>\gamma$. }

\vspace{0.1in}
\noindent
{\sc Proof.} Now, let us write $\lambda=\psi(0,\tilde{\lambda})$, so
$$\alpha=\P_{\theta_0}(\psi(Q,T)>\psi(0,\tilde{\lambda}))\geq \P_{\theta_0}(\psi(0,T)>\psi(0,\tilde{\lambda}))=\P_{\theta_0}(T>\tilde{\lambda}),$$
due to the monotonicity. Hence,
$\gamma\leq \tilde{\lambda}$. Now
$$\P_\theta(\psi(Q,T)>\psi(0,\tilde{\lambda}))\leq \P_\theta(\psi(Q,T)>\psi(Q,\tilde{\lambda}))=P_\theta(T>\tilde{\lambda})$$
and since $\gamma\leq \tilde{\lambda}$, we have
$$P_\theta(T>\tilde{\lambda})\leq P_\theta(T>\gamma),$$
completing the proof. \hfill $\square$

\vspace{0.1in}
\noindent
Hence, the test involving $Q$ and $T$ is uniformly worse than the one involving just $T$. This is because $\psi$ is decreasing in $Q$ whereas $\psi$ is increasing in $T$.  This result has significant implications for the subjective Bayes factor test and using non-informative priors on the nuisance parameters. We note here that it is possible to show Theorem 4 applies to all the nuisance parameter examples appearing in sections 2, 3 and 4.

\vspace{0.2in}
\noindent
{\bf 6. Implications of the results.} This section considers an implication of the result discussed in section 2.1. In particular, we show that a measure of the strength of the evidence in the Bayes factor in favor of the alternative hypothesis in a one-sided testing problem should depend on the sample size and that a single scale independent of the sample size is not always an appropriate one to use. Similar comments apply to the other tests discussed in sections 2, 3 and 4. 

Consider $(X_i)_{i=1}^n$ from a normal distrbution with unknown mean $\theta$ and known variance $\sigma^2=1$ and a test of
$$H_0:\theta=0\quad \mbox{vs} \quad H_1:\theta>0.$$
The classical test for this problem is to reject $H_0$ if $\bar{X}>\gamma$ where $\bar{X}$ is the sample mean and $\gamma$ is chosen so that $P_{\theta=0}(\bar{X}>\gamma)=\alpha$.

The Bayes factor for this problem is given in (\ref{eq33}). Since $\bar{X}=T/n$, the Bayes factor is a monotone increasing function of $\bar{X}$ for any prior $\pi(\theta)$. Then setting
$$ \lambda=\int_{\theta>0} \exp\{n\theta \gamma - \half n\theta^2\}\,\pi(\d\theta)$$
we have $B(\bar{X})>\lambda$ if and only if $\bar{X}>\gamma$ for any prior $\pi(\theta)$, and the power function for both tests is $\beta(\theta)=\P_{\theta}(\bar{X}>\gamma)=\P_{\theta}[B(\bar{X})>\lambda]$. If the true value of $\theta=0$, then $\bar{X}=O(n^{-\half})$ and therefore $B(\bar{X})=O(n^{-\half})$. Hence, $\lambda=cn^{-\half}$ for some constant $c$.

This contradicts the ad-hoc scale introduced by Kass and Raftery (1995), which is
$$
\begin{array}{ll}
B & \mbox{Evidence for alternative hypothesis} \\
1-3 & \mbox{Not worth a mention} \\
3-20 & \mbox{Positive} \\
20-150 & \mbox{Strong} \\
> 150  & \mbox{Very strong.}
\end{array}
$$
The reason is that for a small $n$, a specific value of $B$ will not represent strong evidence for the alternative hypothesis (i.e. for small $n$, if $\theta=0$ this value of $B$ can be reasonably attributed to random chance) while for large $n$ the same value of $B$ will be very unlikely to occur if $\theta=0$ and therefore provides strong evidence in favor of the alternative.

The conclusion is that a Bayes factor can be difficult to interpret in a specific problem and guidance from the classical test in determining strength of evidence will be useful, if not essential.

Selecting $\gamma$ up front to determine a type I error  means that $\lambda$ and $\pi$ are connected and this might seem unreasonable. In fact it is highly reasonable as we now demonstrate. Continuing the example discussed above, suppose the prior for $\theta$ is given by $N(\theta|0,1/\tau)$. Then the Bayes factor is 
$$B(n,\tau)=\sqrt{\frac{\tau}{\tau+n}}\,\exp\{\half\,n^2\bar{X}^2/(n+\tau)\}.$$
Therefore, for the Bayesian test to coincide with the classical test, the corresponding value of $\lambda$ must be a function of both $n$ and $\tau$. Even without considering the equivalency between the Bayesian and classical tests, the chosen $\lambda$ must depend on the value of $\tau$. For example, suppose $n\bar{X}^2=10$, in which case it is reasonable to reject $H_0$. If $n/\tau=10000$, then $B(n,\tau)=1.5$ and according to the above scale the evidence in favor of the alternative is rated as ``Not worth a mention". However, if $n/\tau=100$, then $B(n,\tau)=14.8$ and the evidence is rated as ``Positive" in favor of the alternative. Consequently, there is no universal $\lambda$ that can be chosen to cover all $(n,\tau)$.

\vspace{0.2in}
\noindent
{\bf 7. Discussion.} In this paper we considered using Bayes factors as a means to do Bayesian hypothesis testing. We will consider only one-sided tests of the type $H_0:\theta=\theta_0$ vs $H_1:\theta>\theta_0$ in this section to keep the discussion as concise as possible. Similar comments also apply to the other tests considered in the paper.

The models we consider for one-sided testing problems rely on writing
$$B(T)=\int g(T,\theta)\,\pi(\d\theta)$$
where $T$ is the classical test statistic and $B(T)$ is a monotone increasing function of $T$. If $g(T,\theta)$ is monotone for all $\theta$ then we achieve this for any $\pi$. If not, then we need to restrict $\pi$ to a particular class to ensure $B(T)$ is monotone. 

Now, for any choice of $(\lambda,\pi)$, where the Bayesian would reject $H_0$ if $B>\lambda$, we can find a $\gamma$ for which
$$\lambda=\int g(\gamma,\theta)\,\pi(\d\theta).$$
Then $B>\lambda$ if and only if $T>\gamma$. No matter what $\gamma$ is, the Bayesian and classical tests are equivalent and have the same type I error $P_{\theta=\theta_0}(T>\gamma)$. It is now in our opinion prudent to ensure $\gamma$ is set so that the type I error is a reasonable value for the classical and Bayesian tests. Our added suggestion is that rather than determine $(\lambda,\pi)$ without regard to the type I error, one should set $\gamma$ to give a benchmark type I error and rely on the notion that for any $\pi$ there exists a $\lambda$ for which this $\gamma$ can be realized.

In any case, for the Bayesian pursuing a hypothesis test through a Bayes factor, for the models we have considered, it is a consequence that the role of $(\lambda,\pi)$ is solely to determine the type I error. 

We have also shown that when nuisance parameters are present it is desirable to put the standard non-informative prior on the nuisance parameter. If not, it can be shown that in the examples we have considered, the Bayesian test is uniformly worse than the classical test. 

\newpage

\vspace{0.2in}
\noindent
{\bf References.}

\begin{description}


\item Aitkin, M. (1991). Posterior Bayes factors. {\sl Journal of the Royal Statistical Society Series B} {\bf 53}, 111-142.

\item Bayarri, M.J., Berger, J.O., Forte, A. \& Garcia-Donato, G. (2012). Criteria for Bayesian model choice with application to variable selection. {\sl Annals of Statistics} {\bf 40}, 1550-1577. 

\item Berger, J.O. \& L.R. Pericchi (1996). The intrinsic Bayes factor for model selection and prediction. {\sl Journal of the American Statistical Association} {\bf 91}, 109-122.

\item Casella, G., Giron, F.J., Martinez, M.L.  \& Moreno, E. (2009). Consistency of Bayesian procedures for variable selection. {\sl Annals of Statistics} {\bf 37}, 1207-1228

\item De Santis, F. \& F. Spezzaferri (1997). Alternative Bayes factors for model selection. {\sl Canadian Journal of Statistics} {\bf 25}, 503-515.

\item Garcia-Donato, G. \& Chen, M-H. (2005). Calibrating Bayes factor under prior predictive distributions. {\sl Statistica Sinica} {\bf 15}, 359-380.

\item Jeffreys, H. (1961). Theory of probability. Oxford Univesrity Press, Oxford, U.K.

\item Johnson, V. (2013). Uniformly most powerful Bayesian tests. Submitted.

\item Johnson, V. \& Rossell, D. (2010). On the use of non-local prior densities in Bayesian hypothesis tests. {\sl Journal of the Royal Statistical Society Series B} {\bf 72}, 143-170.

\item Kass, R.E. \& Raftery, A.E. (1995). Bayes factors. {\sl Journal of the American Statistical Association} {\bf 90}, 773-795.

\item Liang, F., Paulo, R., Molina, G., Clyde, M. A. \& Berger, J. O. (2008). Mixtures of $g$-priors for Bayesian variable selection. {\sl Journal of the American Statistical Association} {\bf 103}, 410-423.


\item O'Hagan, A. (1995). Fractional Bayes factors for model comparison. {\sl Journal of the Royal Statistical Society Series B} {\bf 57}, 99-138.

\item Schwarz, G.E. (1978). Estimating the dimension of a model. {\sl Annals of Statistics} {\bf 6}, 461-464.

\item Shi, N. \& Tao, J. (2008). Statistical Hypothesis Testing, Singapore, World Scientific Publishing Co.


\end{description}

\end{document}